\documentclass[a4paper]{PoS}

\usepackage{subfigure}
\usepackage{epsfig}

\title{The global electroweak f\/it and constraints on new physics with Gf\/itter}

\ShortTitle{The global electroweak f\/it and constraints on new physics with Gf\/itter}

\author{\speaker{D\"orthe Ludwig} for the Gf\/itter group\\
        DESY Hamburg, Unversity of Hamburg\\
        E-mail: \email{doerthe.ludwig@desy.de}}

      \abstract{ The thourough investigation of radiative corrections
        allows to gain information on physics processes at higher
        energy scales than those directly accessible by current
        experiments. As a consequence, using electroweak precision
        measurements in conjunction with state-of-the-art SM
        predictions e.g. allows the estimation of a preferred mass
        range for the SM Higgs boson mass. Physics beyond the Standard
        Model can modify the relations between electroweak observables
        and their theoretical predictions. Such effects can be
        parametrized in terms of effective, so-called oblique
        parameters. A global f\/it of the electroweak SM, as performed
        with the Gf\/itter package~\cite{ref:Gfittermain}, allows the
        determination of the oblique parameters and to probe physics
        models and to set constraints on their free parameters.

        In this paper we present updated results of the global
        electroweak SM f\/it taking into account the latest
        experimental precision measurements and the results of direct
        Higgs searches from LEP and Tevatron.  Through the formalism
        of oblique parameters we obtain constraints on BSM models with
        universal and warped extra dimensions.  In constrast, taking
        into account heavy flavor observables, (g-2)$_\mu$, and dark
        matter predictions allows to constrain the parameter space of
        the minimal supergravity model (mSugra).}

      \FullConference{35th International Conference of High Energy
        Physics - ICHEP2010,\\July 22-28, 2010\\Paris France}

\begin{document}

\section{The global electroweak fit }

\begin{figure}
\centering
\includegraphics[width=0.49\textwidth]{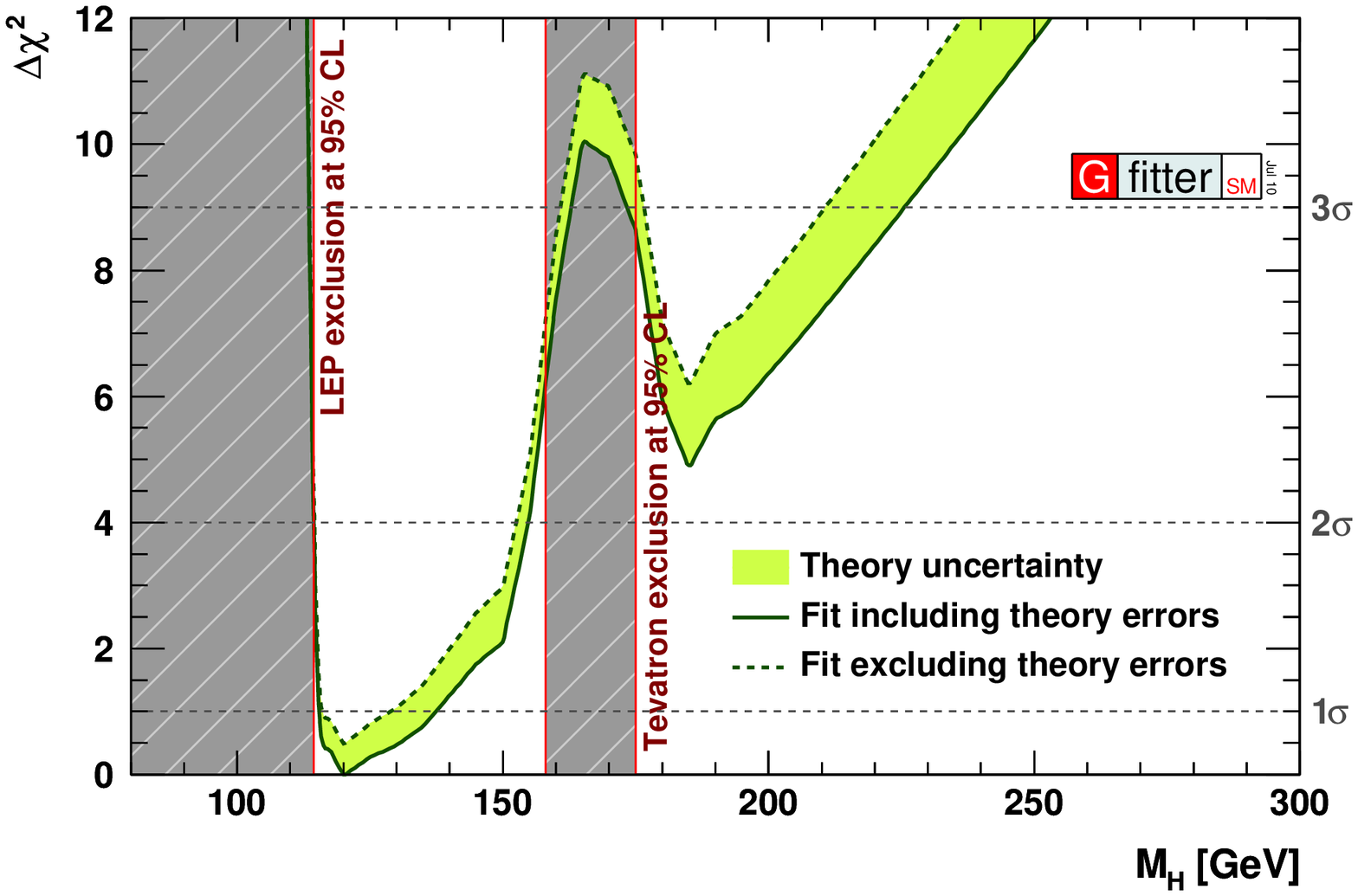}
\includegraphics[width=0.49\textwidth]{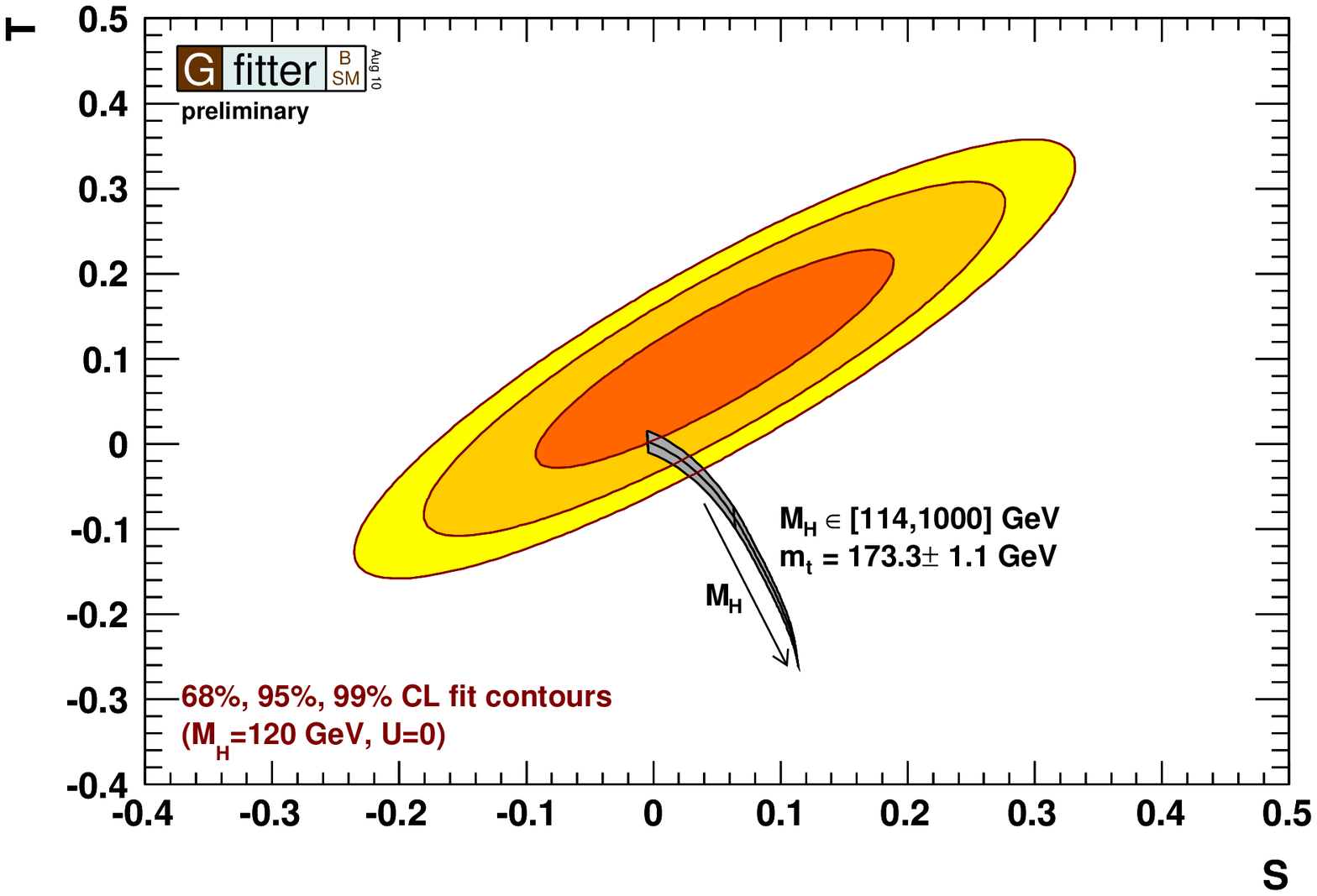}
\caption[]{Left: $\Delta\chi^2$ profile as a function of $M_H$ for
  the electroweak fit including results of direct Higgs searches
  at LEP and Tevatron (shaded areas). Right: Fit result of the oblique
  parameters. Shown are the 68\%, 95\%, and 99\% CL allowed regions in
  the $S$-$T$-plane with $U=0$ for a reference SM with $M_H=120$\,
  GeV and $m_t=173.3$\,GeV. The grey area illustrates the SM
  prediction for various values of $M_H$ and $m_t$.}
  \label{fig:Higgs_SvsT}
\end{figure}
In the global electroweak fit predictions for precision observables
are compared with the most recent measurements done by LEP, SLC, and
Tevatron. A detailed list of all data used in the fit can be found
in~\cite{ref:Gfittermain}. The free floating parameters are $M_Z,
M_H, m_t, m_b, m_c, \Delta\alpha^{(5)}_{\rm had}$, and
$\alpha_S(M_Z^2)$ where the latest average for the $m_t$ as well as
the newly obtained exclusion limits for $M_H$~\cite{ref:TopHiggsmass}
have been used.

The fit converges at a $\chi^2$ minimum of $16.4(17.8)$ excluding
(including) the direct Higgs searches. The respective p-values based
on toy Monte Carlo experiments are 0.23 (0.22) where no individual
pull value exceeds $3\sigma$. One of the most important results of the
electroweak fit is the estimation of Higgs mass. The $\chi^2_{min}$ is
found at $M_H = 84.2^{+30.3}_{-23.3}$\,GeV ($M_H =
120.6^{+17.0}_{-5.2}$\,GeV) with a $2\sigma$ interval of
[40.3,159.2]~GeV ([114.4, 154.9]~GeV).
Figure~\ref{fig:Higgs_SvsT}~(left) shows the $\Delta\chi^2$ profile as
a function of $M_H$ for the fit including the direct Higgs searches.
The increase of $\Delta\chi^2$ at the 95\%\, CL exclusion limits from
LEP and Tevatron (shaded areas) is clearly visible.

\section{The oblique parameters and constraints on beyond the SM
  physics models}

The main assumption that led to the introduction of the oblique
paramters~\cite{ref:STU} is that high-scale BSM physics appears only
through vacuum polarisation corrections. The electroweak fit is
sensitive to BSM physics through these oblique corrections which can
be described through the $STU$ parametrization: $O_{meas} =
O_{SM;ref}(M_H;m_t )+c_SS+c_TT +c_UU$. The $STU$ parameters measure
deviations from electroweak radiative correction that are expected in
the reference SM determined by the chosen $m_t$ and $M_H$.

In our analysis the SM reference point is chosen to be at
$M_H=120$\,GeV and $\rm m_t=173.3$\,GeV. $S, T$, and $U$ are derived
from a f\/it to electroweak observables and are compatible with 0 as
illustrated in Fig.~\ref{fig:Higgs_SvsT}~(right) showing the 68\%,
95\%, and 99\% CL level allowed regions in the $S$-$T$-plane for
$U=0$. The grey area shows the SM prediction highlighting the
logarithmical dependence of $S$ and $T$ on $M_H$. Small values of
$M_H$ are compatible with data.

\subsection*{Extra dimension models}

\begin{figure}[t]
\centering
\includegraphics[width=0.49\textwidth]{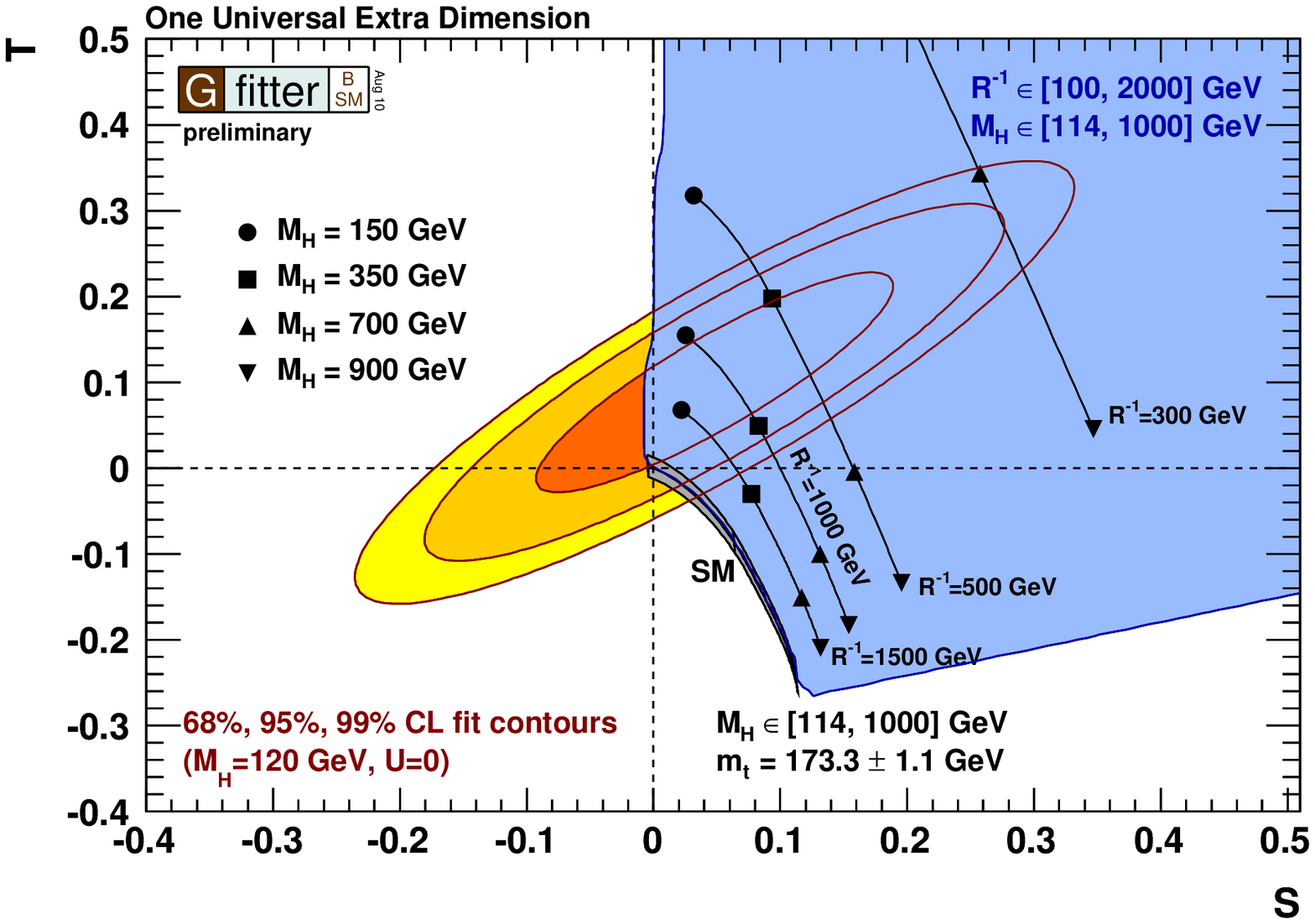}
\includegraphics[width=0.49\textwidth]{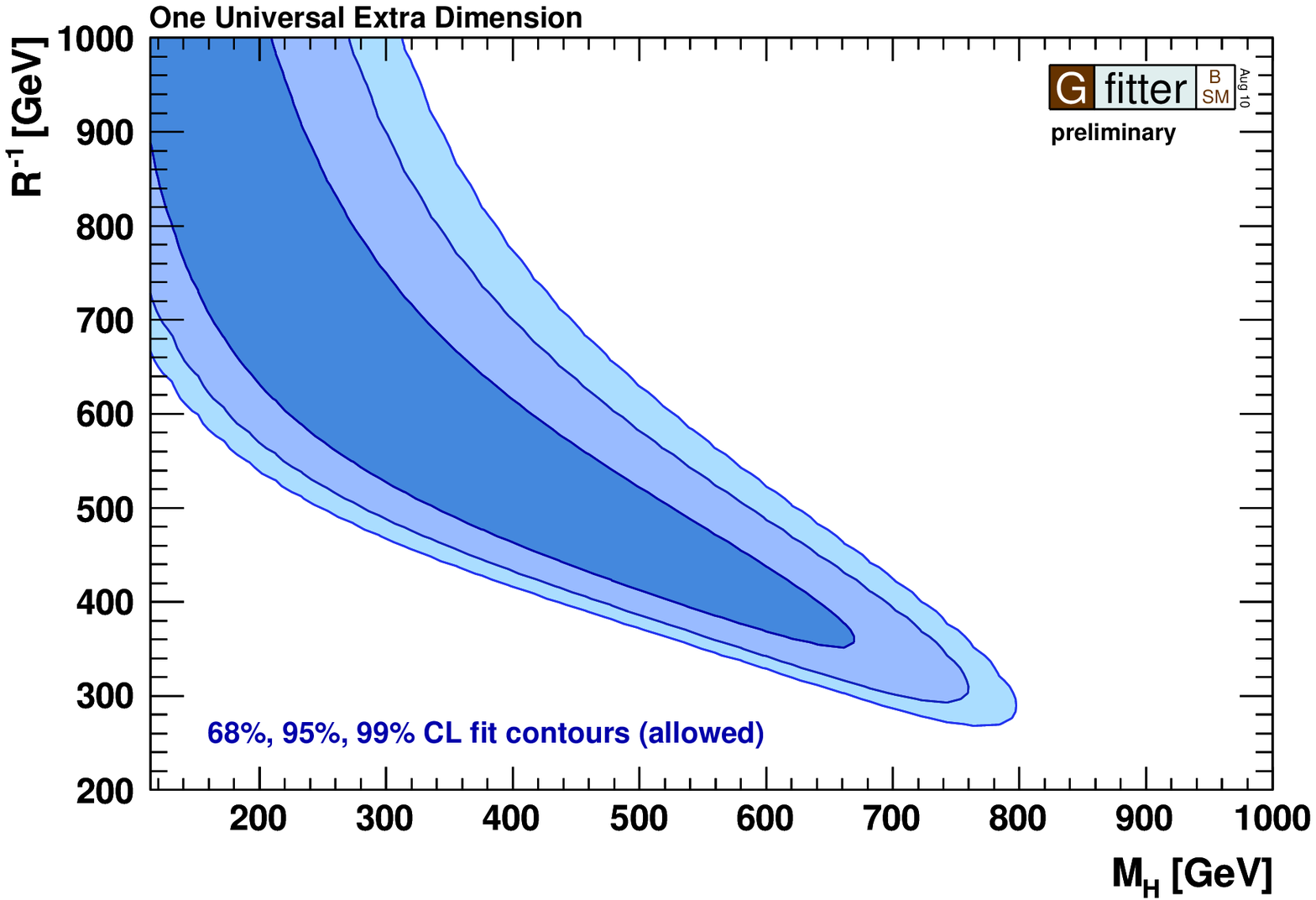}
\caption[]{Left: Comparison in the S-T-plane between CL contours from
  f\/its to the electroweak precision data and predictions of the
  UED model. For illustration some benchmark points are depicted. With
  increasing compactif\/ication scale the ST values converge to the SM
  predictions. Right: Contours of 68\%, 95\% and 99\% CL obtained from
  scans of f\/its with f\/ixed variable pairs $R^{-1}$ and $\rm
  M_H$.}
  \label{fig:UED}
\end{figure}
The universal extra dimension model (UED) ~\cite{ref:UED} allows all
SM particles to propagate into the extra dimensions. Their
compactif\/ication leads to Kaluza-Klein (KK) modes.  The free
parameters of this model are the number of extra dimensions $d_{\rm
  UED}$ which is f\/ixed to one in our analysis and the
compactif\/ication scale $R^{-1}$ where $R$ is the size of the extra
dimension.  The $S$, $T$, and $U$ parameters mainly depend $R^{-1}$,
$M_H$, and $m_t$. The result is illustrated in
Fig.~\ref{fig:UED}~(left). The blue region indicates the
$S$-$T$-parameter space allowed for various parameter sets in the UED
model. For large compactif\/ication it approaches the SM prediction.
Only small $M_H$ are allowed. If the compactif\/ication becomes
smaller the BSM contribution needs to be compensated by larger $M_H$.
A comparison with electroweak precision data allows the exclusion of
$R^{-1} < 300$\,GeV and $M_H > 800$\,GeV as can be seen in
Fig.~\ref{fig:UED}(right) in which the 68\%, 95\%, and 99\% CL allowed
regions are shown.

A different higher-dimensional approach has been proposed by Randall
and Sundrum\cite{ref:WEDI}. The geometry of this model is
characterized by one warped extra dimension conf\/ined by two
three-branes, one of them containing the SM f\/ields. The generation
of the weak scale from a larger scale is achieved by introducing a
warp factor altering the known four-dimensional metric. The inverse
of the warp factor $L$ and the Kaluza-Klein scale $M_{\rm KK}$ are
free model parameters. In a slightly extended version of the original
minimal model SM gauge bosons and fermions can propagate in the
five-dimensional warped bulk region~\cite{ref:WEDII}.
Figure~\ref{fig:WED_SU}~(left) shows again in blue the allowed region
in the $S$-$T$-plane. For large values of $L$ high $M_{\rm KK}$ values
are preferred by the data.  However, the higher $M_H$ the lower
$M_{\rm KK}$ values can be compensated.

\section{Minimal Supergravity}

\begin{figure}[t]
\centering
\includegraphics[width=0.49\textwidth]{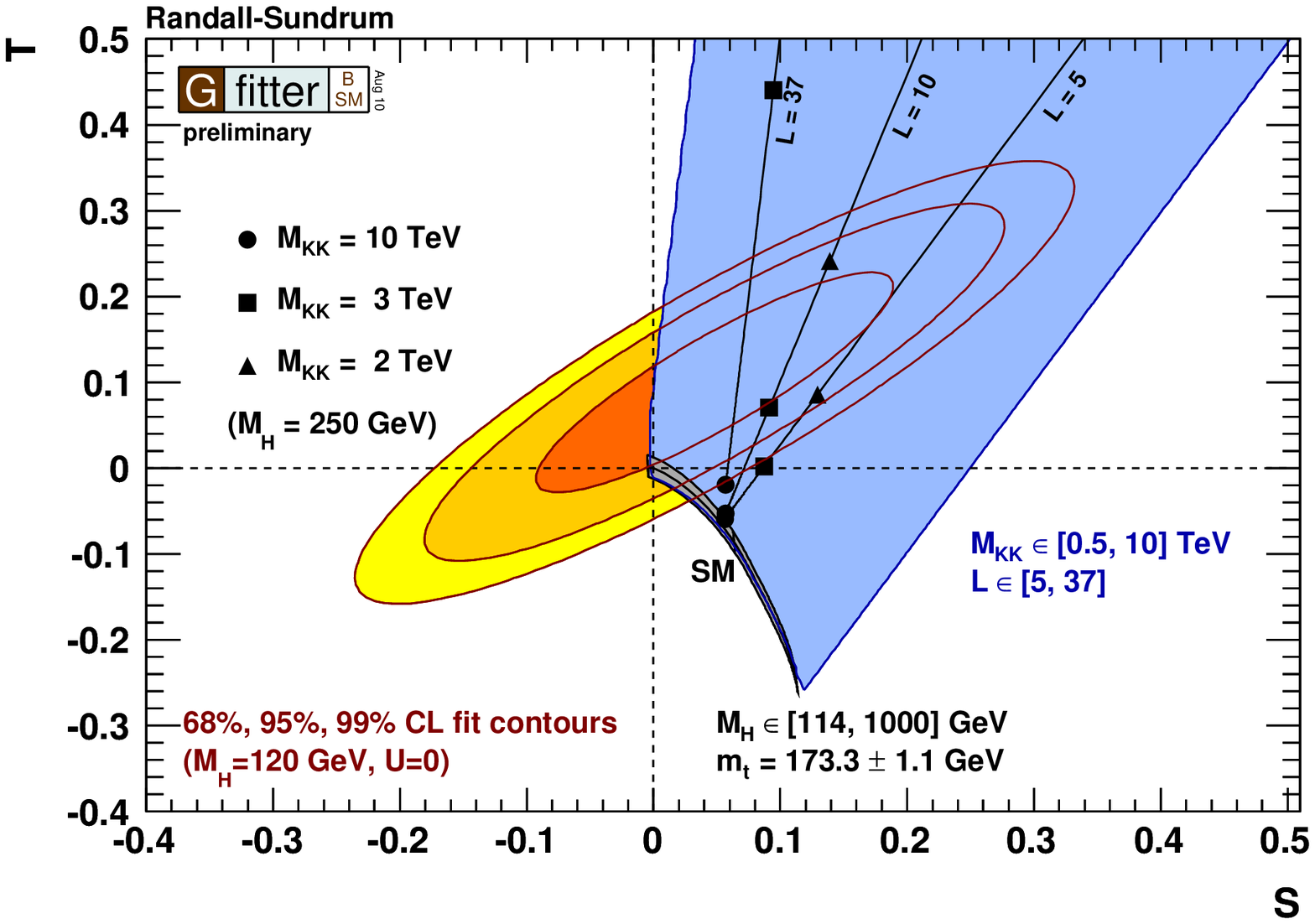}
\includegraphics[width=0.49\textwidth]{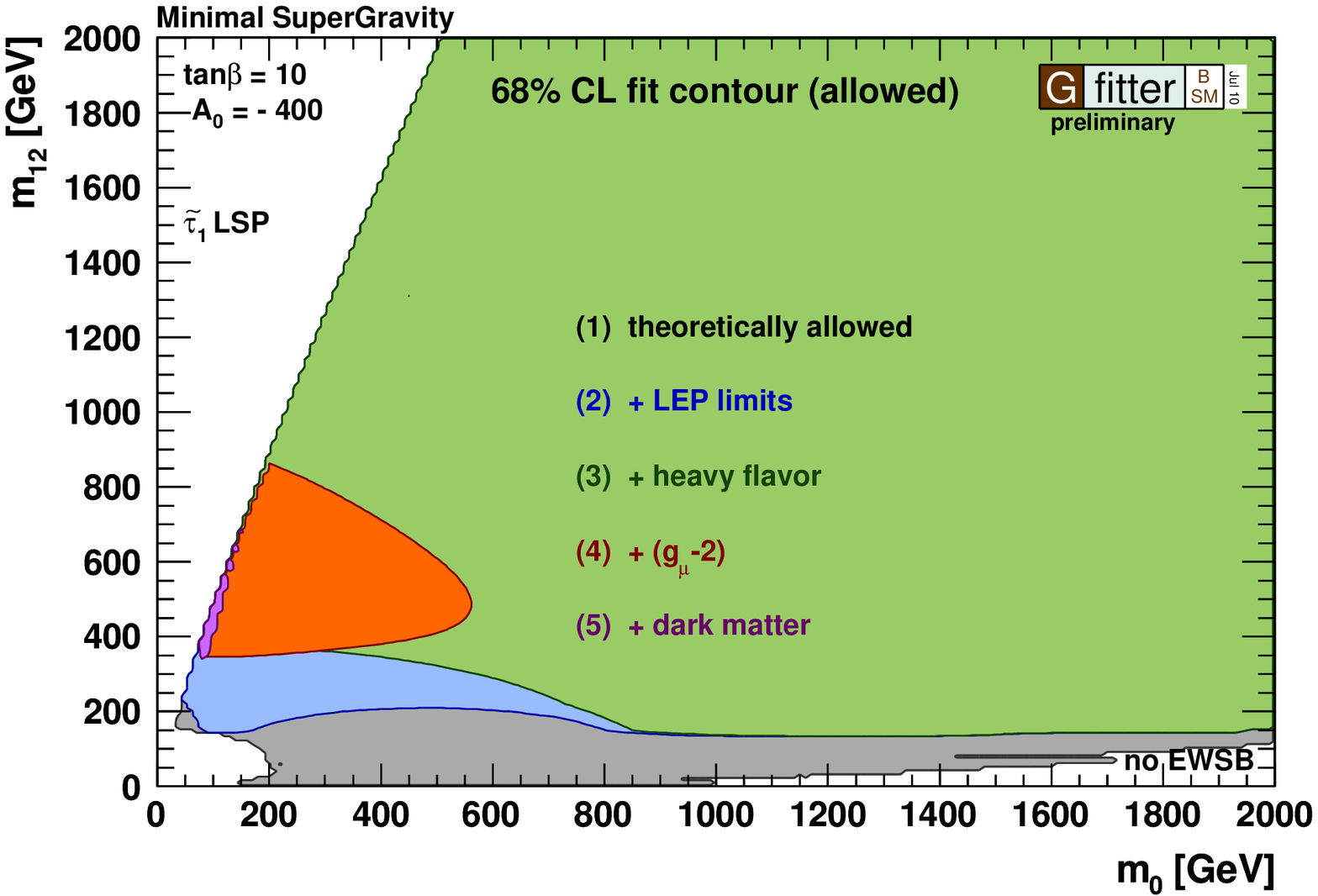}
\caption[]{Left: Comparison in the S-T-plane between CL contours from
  f\/its to the electroweak precision data and predictions from models
  with one warped extra dimensions. For illustration several benchmark
  points are depicted. Right: Contours of 68\% CL obtained from scans
  with f\/ixed $m_0-m_{12}$ pairs taking into account constraints
  from (2) LEP (blue), (3)heavy flavor (green), (4) $(g-2)_{\mu}$
  (orange), and (5) dark matter (purple). }
  \label{fig:WED_SU}
\end{figure}

Minimal supergravity is one of the most investigated supersymmetry
models by current collider experiments. One highly constraining
breaking mechanism suggests that the breaking is mediated by the
gravitational interaction. Some of the free parameters of this model
are $m_{0(12)}$, the mass of scalar particles (fermions) at the GUT
scale, and $\tan\beta$, the ratio of the two Higgs vacuum expectation
values. Figure~\ref{fig:WED_SU}~(right) shows the constraints on $m_0$
and $m_{12}$. Whereas the impact of the LEP and heavy flavor data do
not severely limit these two parameters the $(g-2)_{\mu}$ and dark
matter measurements clearly favor small values.

\section{Conclusion}

The global electroweak f\/it shows an excellent agreement of the SM
with data. Including the latest experimental results from Tevatron for
$m_t$, $M_W$, and $M_H$ results in a $\chi^2_{min}$ at $M_H =
120.6^{+17.0}_{-5.2}$\,GeV. However, this result may change if BSM
physics is present. For many new physics model contributions from a
larger Higgs mass could be compensated and be still compatible with
current data. A more detailed list including the Littlest Higgs model,
warped extra dimensions with custodial symmetry and a fourth
generation model can be found in~\cite{ref:Gfittermain}. As the LHC and
Tevatron contribute further to the electroweak precision measurements
tighter constraints might be set on many BSM models.  Therefore, a
continuous development of the electroweak fit and the oblique
parameters will be carried out.

\end{document}